\documentstyle[11pt,tss_paspconf,epsf]{article}

\begin{document}

\title{The Dwarf Spheroidal Galaxies in the Galactic Halo}

\author{G. S. Da Costa}
\affil{Research School of Astronomy \& Astrophysics, Institute of Advanced
Studies, The Australian National University}

\begin{abstract}
In the first part of this contribution the observed velocity 
dispersions in Galactic halo dwarf spheroidal (dSph) galaxies are reviewed, 
and the consequences for dark matter content outlined.  The results are
suggestive of a dSph dark matter mass of $\sim$2$\times10^{7}$ solar masses, 
independent of the luminosity of the dSph.  Alternatives to the dark matter 
interpretation are also briefly discussed.  In the second part of the 
contribution the emphasis is on the stellar populations of Galactic dSphs.  
Recent results for the ages of the oldest populations are presented.  These 
data, together with similar recent results for Galactic
halo and LMC globular clusters, indicate that regardless of the subsequent
star formation history, {\it the initial epoch of star formation was well
synchronized throughout the entire proto-Galactic halo}.  The implications of 
the first high dispersion studies of element abundance ratios in Galactic dSph 
red giants are also discussed.  For the Sagittarius dSph in particular, the 
observed abundance ratios show good agreement with expectations for an episodic 
star formation history, and such a history is in fact deduced 
from colour-magnitude diagram studies.

\end{abstract}


\section{Introduction}

At the present time, the Galaxy has nine identified dwarf spheroidal (dSph) 
galaxy companions: Sculptor, Fornax, Draco, Ursa Minor, Leo~I, Leo~II, Carina,
Sextans and Sagittarius.  These systems appear to be typical members of the
dSph/dE class of low luminosity, small, low surface brightness galaxies. 
The wording ``at the present time'' is deliberately chosen because it is 
unclear whether the current population of Galactic dSph companions represents 
most, or only a small fraction, of the dSphs that existed in the halo at early 
times in the life of the Galaxy.  For example, the Sgr dSph is currently 
undergoing a significant interaction with the Milky Way and, within the next 
billion years or so, it probably will no longer be recognizable as a distinct 
object, having merged into the Galactic halo.  Thus, as other contributions 
in this volume will emphasize, the disruption of dSph galaxies might have 
contributed significantly to the make-up of the Galactic halo (see also Mateo 
1996).  In this review, however, I will concentrate on the properties of the 
current Galactic halo dSphs.  Indeed, these galaxies are particularly relevant 
to the conference theme of ``Bright Stars \& Dark Matter'', since they are the 
only Galactic halo systems where both stars and dark matter are found together 
in bound systems.  

In the first part of this contribution, then, the observed velocity dispersions
of the Galactic dSphs and the implications for dark matter contents
will be considered.  This is followed in the second part by a discussion
of some new results derived from characteristics of dSph stars.

\section{Galactic Dwarf Spheroidal Velocity Disperions}

In 1983 the late Marc Aaronson published a paper (Aaronson 1983) that
revolutionized our concept of the masses and dynamics of dSph galaxies.
His paper indicated that the velocity dispersion of the Draco dSph, albeit
based on only 5 observations of 4 stars, was at least $\sim$6.5 kms$^{-1}$
and that, as a consequence, the mass-to-light ratio of this dSph exceeded
that of globular clusters by at least an order of magnitude.  Since the
stellar population of Draco is apparently similar to those of Galactic 
globular clusters, Aaronson's result implied the existence of a substantial
amount of dark matter in this dSph.  Since the publication of that paper
there have been a number of similar studies of Draco and of the other 
Galactic dSphs, with increasingly large samples of stars.  Yet the basic
result has remained the same.  Indeed, with the publication of results for
Leo~I (Mateo et al.\ 1998), we can now say that {\it all} the Galactic
dSphs appear to contain significant amounts of dark matter -- see 
Mateo (1997) and Olszewski (1998) for recent reviews of this subject.

In this section the recent work on Leo~I (Mateo et al.\ 1998) is first
considered. Other than the large Galactocentric distance, 
which makes the target
stars relatively faint (and therefore required use of the Keck telescope
for the observations), the Mateo et al.\ (1998) study of the velocity dispersion
of Leo~I is typical of existing Galactic dSph velocity dispersion studies.  It
therefore provides an example with which to highlight the steps (and the
potential pitfalls) in the process by which observations of individual
radial velocities are transformed into a dSph mass-to-light ratio estimate.  
The dark matter content implications of these results, and those for other 
Galactic dSphs, are then presented.  The section concludes with a brief
discussion of an alternative interpretation of the large observed velocity
dispersions -- that the dSphs are undergoing tidal disruption and are thus
not in virial equilibrium.

\subsection{Leo I -- A Case Study}

The sample of Leo~I stars observed by Mateo et al.\ (1998) consists of 33
red giants selected from a colour-magnitude study.  These stars were observed
at high dispersion (R $\approx$ 34,000) but the resultant spectra have
relatively low S/N ratio.
Velocities are obtained by cross-correlating the spectra with high S/N
spectra of radial velocity standards.  The high systemic velocity
of Leo~I, $\sim$290 kms$^{-1}$, assures us of Leo~I membership for all
the candidates observed, though for some other Galactic dSphs
this member/non-member discrimination is not as clear cut.  A total of 40 individual measurements were made with
the typical velocity error being $\sim$2.2 kms$^{-1}$ (the actual errors range
from 1.4 to 4.8 kms$^{-1}$ depending on the S/N ratio of the spectrum).  Then,
based on a number of different techniques (e.g.\ weighted standard deviation,
bi-weight estimator, maximum likelyhood), all of which produce similar
values, the observed velocity dispersion for this sample
of Leo~I stars is $\sigma_{obs}$ = 8.8 $\pm$ 1.3 kms$^{-1}$.  Note that the
error associated with this dispersion comes principally from the 
``sampling error'' arising from the finite size of the observed sample, and that
the value applies to the core of Leo~I, since
all but one of the stars observed have (in projection at least) radial
distances less than the core radius.  Further, within this limited radial 
range, there is no indication of any change in $\sigma_{obs}$ with location,
nor is there any evidence for systematic rotation of
Leo~I, at least in the core region.  These latter results are also 
commonly found for other Galactic dSph systems.

Mateo et al.\ (1998) then apply what is now standard formalism to calculate
from the observed dispersion a {\it central} mass-to-light ratio 
($\rho_{0}$/I$_{0,V}$), using the observed central surface brightness 
of Leo~I, and a {\it total} mass-to-light ratio (M/L$_{total,V}$)
from the total integrated magnitude of the dSph.  Both these
calculations require a length scale; the core radius derived from the 
observed surface brightness (or surface density) profile is usually used.
These structural parameters are now (at least
moderately) well known for all Galactic dSphs, though they are, of course,
based on the stellar distribution which may not reflect the underlying
mass distribution.  The scale factors required in these calculations are
usually taken as those for the King (1966) model which best fits the 
surface brightness/density profile.  These models are appropriate for
spherically symmetric systems with isotropic velocity distributions whereas
the Galactic dSphs have significant flattening yet lack systematic rotation;
consequently, they presumably have anisotropic velocity distributions.
The use of King model parameters, however, is not regarded as crucial
(e.g.\ Merritt 1988); much more fundamental (e.g.\ Pyror \& Kormendy 1990,
Pyror 1994)
is the implicit assumption here that ``mass follows light'', an assumption
for which there is little justification at present.

Applying this formalism, Mateo et al.\ (1998) find $\rho_{0}$/I$_{0,V}$ =
3.5 $\pm$ 1.4 and M/L$_{total,V}$ = 5.6 $\pm$ 2.1 indicating that M/L$_{V}$
for Leo~I could lie anywhere between $\sim$2 and 8 in solar units.  
How then are we to
interpret these values?  Mateo et al.\ (1998) point out that data
for low central concentration globular clusters (e.g.\ Pryor \& Meylan 1993), 
which are the ones for which mass segregation effects should be minor, yield
$<$(M/L$_{V}$)$>$ = 1.5 $\pm$ 0.1 when analyzed in
the same way as the Leo~I observations.  At first sight a direct comparison
of this mean value with the derived M/L values for Leo~I doesn't convincingly
argue for the presence of a significant dark matter content in the dSph.
However, we must keep in mind that the stellar population of Leo~I is not 
that of a globular cluster (as is the case for many of the Galactic dSphs).
In fact, Lee et al.\ (1993) have shown that the stellar population of Leo~I
is dominated by stars of intermediate-age (i.e.\ ages $\sim$ 2 -- 10 Gyr)
so that the ``mean age'' of Leo~I is considerably younger than that of the
globular clusters.  Mateo et al.\ (1998) have constructed simple stellar
population models to correct for this effect and have calculated that
for a valid comparison with the Galactic globular clusters, the observed 
Leo~I M/L$_{V}$ value should be increased by a factor of approximately two.
In other words, after compensating for stellar population differences, we
have (M/L$_{V}$)$_{Leo I}$ $\approx$ 9 and (M/L$_{V}$)$_{glob~cl}$ $\approx$ 
1.5 -- a clear indication that there is a significant dark matter component
in Leo~I.

One might question the strength of this conclusion on the basis that the
Mateo et al.\ (1998) data, like the situation for many of the
Galactic dSphs, are essentially single epoch observations and that as a
result, the presence of binary stars might have inflated the observed 
dispersion and caused the M/L ratio to be overestimated.  The extensive work 
of Olszewski et al.\ (1996), however, has shown that this is not a valid objection.  These authors have extensive repeat observations, extending over
many years, of a large number of stars in the Galactic dSphs Draco and
Ursa Minor.  Indeed, despite the fact that the minimum possible binary period,
given the radii of the red giants observed in these programs, is approximately
six months, Olszewski et al.\ (1996) have sufficient data to investigate the
binary frequency in these dSphs.  Among the stars observed, they find six 
likely binaries, four in Ursa Minor and two in Draco.  Then, via an exhaustive
set of simulations, Olszewski et al.\ (1996) convert this observed binary
frequency among red giants into an estimate of the overall binary star
frequency in the dSphs.  Surprisingly, they find that the binary frequency
in Draco and Ursa Minor might be as much as three times higher than it is
in Population I samples, and as much as five times higher than is the case
for field Galactic halo samples.  Discussion of this intriguing result is
beyond the scope of this contribution.  Nevertheless, the extensive simulations
of Olszewski et al.\ (1996) show that even with such a high binary frequency,
the effect of undetected binaries on the velocity dispersion determined from
datasets similar to that of Mateo et al.\ (1998) for Leo~I, is small and
cannot be used as an explanation for the high mass-to-light ratio.  

\subsection{The Dark Matter Interpretation}

The observed mass-to-light ratios for the Galactic dSph 
galaxies\footnote{The Sgr dSph 
is excluded from the discussion here as the extent to which its obvious
interaction with the Milky Way compromises the interpretation of the
observed velocity dispersion measurements is unclear (but see also 
Ibata et al.\ 1997).} 
range from values in excess of 50 for the low
luminosity systems Draco and Ursa Minor (e.g.\ Armandroff et al.\ 1995) to
values of order 5 for the more luminous systems (e.g.\ Mateo 1998 and the
references therein).  As noted above, the star formation histories of
the Galactic dSphs vary considerably from system to system and a proper
comparison of M/L values must then take this into account.  We follow the procedures of Mateo et al.\ (1998) and reduce the observed M/L values (taken
from Mateo 1998)
to those which would be expected if the dSphs were composed only of stars
similar to those in globular clusters.  The results of this process are
shown in Fig.\ 1.  As has been noted many times in the past (using uncorrected
values), Fig.\ 1 reveals a general correlation with the least luminous systems 
showing the highest M/L values.  The dashed curve in Fig.\ 1 is the 
relation (cf.\ Mateo et al.\ 1998) expected if the luminous stars in each 
dSph are embedded in a dark matter halo of constant mass, independent of
the luminosity of the dSph.  That is, the
line is derived from the relation:
\begin{center}
(M/L)$_{total,V,corr}$ = (M/L)$_{V, stars}$ + M(Dark Matter)/L$_{stars, V, corr}$
\end{center}
where, since we have corrected to a globular cluster like population,
(M/L)$_{V, stars}$ = 1.5 in solar units.  For the curve shown in Fig.\ 1,
M(Dark Matter) = $2\times10^{7}$ solar masses.

\begin{figure}
\begin{center}
\begin{minipage}{20cm}
\epsfxsize=9.7cm
\epsfbox {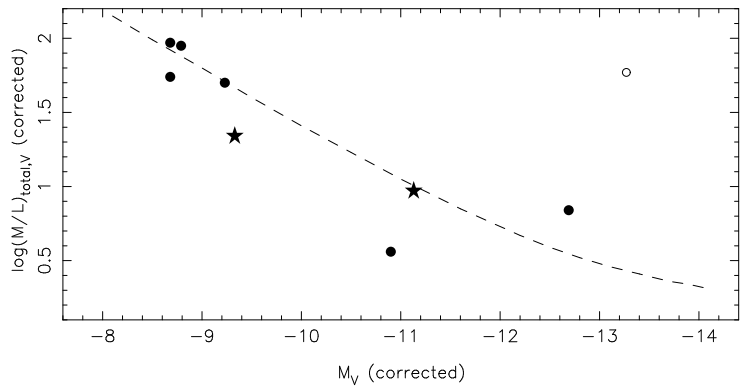}
\end{minipage}
\caption{A plot of the logarithm of the total (as distinct from central)
visual mass-to-light ratio against absolute visual magnitude for the 
Galactic dSphs.  Both the M/L and M$_{V}$ values, taken from Mateo (1998),
have been corrected for the
stellar population differences between the dSphs, as outlined by Mateo et al.\
(1998).  The Leo dSphs, which have Galactocentric distances beyond 200 kpc,
are plotted as star-symbols, while the Sgr dSph, which appears to be strongly perturbed by its
interaction with the Milky Way, is plotted as an open symbol.  The dashed line
is the relation M/L = 1.5 + M$_{DM}$/L with M$_{DM}$ = $2\times10^{7}$ solar
masses.}
\end{center}
\end{figure}

The constant dark matter mass line gives a reasonable representation
of the Galactic dSph points in Fig.\ 1, but we should not take this concordance
too seriously.  Recall that the mass estimates use scale
factors from spherically symmetric isotropic King (1966) models, which are
unlikely to be appropriate for real (flattened, anisotropic) dSphs. More significantly, the mass estimates are based
on the assumption that ``mass follows light''.  This latter assumption 
can be investigated (e.g.\ Pyror 1994) if radial velocities are determined 
for large samples ($\ge$100 stars)
of dSph stars in regions that extend well beyond the core radius of the
visible population.  From such samples
the observed velocity dispersion profile can be constructed and compared
to the predictions of models which either retain or relax the ``mass follows
light'' assumption.   Such datasets are just now becoming available.  In all 
cases the observed velocity dispersion
profiles are flatter than the predictions based on the King model
that best fits the observed surface density profile (e.g.\ Fig.\ 1 of Mateo 
1997), indicating a mass distribution more extended than the light.  This
is an area where we can expect to see interesting new developments in the
near future, but it seems unlikely that the new models and new data will
remove the requirement for a significant dark matter content in the
Galactic dSph galaxies.

\subsection{Alternatives to the Dark Matter Interpretation?}

The oft-discussed alternative to the ``significant dark matter content''
interpretation for the large observed velocity dispersions in Galactic dSphs
is that the dSphs are, in fact, tidally disrupted remnants that are not
in virial equilibrium (e.g.\ Kroupa 1997 and references therein).  
Consequently, the observed velocity dispersions cannot be true 
reflections of the actual dSph masses.  There is insufficient space here
to discuss this alternative view in detail.  However, when
considering its validity, at least
two points should be kept in mind.  First, the Galactic dSphs exhibit
correlations between quantities such as luminosity, surface brightness,
length scale and mean metal abundance.  While Kroupa (1997) indicates how 
a surface brightness -- absolute magnitude correlation might be expected among 
a set of tidally disrupted remnants, the corresponding absolute magnitude --
mean abundance and surface brightness -- mean abundance correlations among
the Galactic dSphs have
no explanation in this scenario.  These correlations, which cover
a luminosity range greater than a factor of 100, are also followed by 
the dSph companions to M31, by the {\it isolated} Local Group dSph Tucana,
and even by dSphs beyond the Local Group (see, for example, Fig.\ 18 of
Caldwell et al.\ 1998).  The similarity of these relationships
in different environments then argues rather strongly against the 
interpretation of the Galactic dSphs as nothing but a set of tidally disrupted 
remnants.  

Second, the Galactic dSphs Leo I and Leo II have Galactocentric distances
that exceed 200~kpc.  Thus neither of these dSphs are
subject to Galactic tides to anything like the same extent as the inner
dSphs (R $\approx$ 65 -- 90 kpc, excluding Sgr).  
Yet neither Leo~I nor Leo~II lies in a
distinctly different location, relative to the other Galactic dSphs, in 
Fig.\ 1, for example.  
Further, both Leo systems have M/L values
that imply the presence of dark matter.  This lack of separation
between the ``near'' and ``far'' Galactic dSphs is then
a further argument against
the tidally disrupted remnants interpretation for the Galactic dSphs.

Nevertheless, if the tidally disrupted scenario doesn't apply to all 
Galactic dSphs, we can at least ask if there are any particular cases
(other than the Sgr dSph, which is clearly being strongly affected by
Galactic tides)
where such a scenario might apply.  The signatures of such a situation 
might well include large scale streaming motions\footnote{Tidal disruption 
models, e.g.\ Piatek \& Pryor (1995) and Oh et al.\ (1995), show that
this process produces large scale ordered motions rather than large
random motions.}, sub-structure and ``extra-tidal'' stars as well as
appreciable line-of-sight depth.  Given these potential signatures it is
then interesting to consider recent results for the Galactic dSph Ursa Minor.
This dSph is one of the closest of the Galaxy's dSph companions and it
has one of the largest apparent M/L values (M/L$_{V}$ = 77 $\pm$ 13, Armandroff
et al.\ 1995).  With an ellipticity e = 0.55, Ursa Minor is also the flattest 
of the Galactic dSph companions.  

The results of interest are as follows:\\
(1) Kleyna et al.\ (1998) have used deep
wide-field CCD imaging to confirm and establish the statistical significance
of an asymmetry in the stellar distribution along the major axis of
Ursa Minor.  In a dynamically stable system such asymmetries should be erased 
on timescales that are of order at most a few crossing times, or $\le$10$^{9}$
years in Ursa Minor.\\  
(2) Kroupa (1997) has generated a tidally disrupted model for Ursa Minor in
which he suggests that the true major axis of the dSph lies at a significant
angle to the plane of the sky.  This generates a line-of-sight depth which,
for a core radius of $\sim$200 pc, Kleyna et al.\ (1998) estimate as
$\sim$0.04 mag in size.  These authors then searched for this effect by
calculating the mean apparent magnitude of samples of horizontal branch stars
along the major axis.  They find that $<V>_{SW} - <V>_{NE}$ = 0.025 $\pm$ 0.021
mag and $<I>_{SW} - <I>_{NE}$ = 0.036 $\pm$ 0.035 mag, 
which is suggestive of the
postulated effect but far from a convincing demonstration.\\
(3) Both Hargreaves et al.\ (1994) and Armandroff et al.\ (1995) have 
reported a velocity gradient approximately along the {\it minor} axis in 
Ursa Minor.  Tidal disruption models generally have streaming motions that are revealed as apparent major-axis rotation, but, given the fact that of all 
the dSphs so far
investigated, Ursa Minor is the only one to show {\it any} rotation
signature (whether about the major or minor axis), it is not unreasonable
to suggest that this observed motion is the consequence of a tidal effect.
The size of these ordered motions ($\sim$3 kms$^{-1}$ per 100 pc
in projection), however, is considerably smaller than the observed dispersion
($\sim$9 kms$^{-1}$).  Thus it is unlikely that the observed M/L for Ursa
Minor is significantly overestimated.

Then, given that Ursa Minor is seemingly the best object for which a tidal 
disruption model might be viable, yet such a model is not 
compellingly required, 
it seems reasonable to conclude that the Galactic dSph galaxies do indeed 
contain significant amounts of dark matter.

\section{Galactic Dwarf Spheroidal Stellar Populations}

One of the most interesting developments in the study of dSph galaxies
in recent years has been the recognition that the star formation histories
vary significantly from dSph to dSph (see, for example, the recent
reviews of Da~Costa 1997a, 1998, Mateo 1998, and the references therein). 
However, rather than consider the latest results on
their evolutionary history (e.g.\ Stetson et al.\ 1998)
two issues relevant to the conference theme
will be considered.  The first is the age of
the oldest stars in the Galactic dSph companions as compared
to the age of the oldest
stars in other Galactic halo objects.  The second is a discussion of how
element abundance ratios in dSph red giants, recently determined for the
first time, compare with similar data for field halo stars.

\subsection{The Age of the Oldest Populations}

All of the Galactic dSphs are known to contain RR Lyrae variable stars.  
Since such variables are also found in Galactic halo globular clusters,
the occurrence of RR Lyraes in dSph galaxies has conventionally been taken
as evidence for the presence of a stellar population in each dSph 
that has an age 
comparable to those of the Galactic globular clusters.  The relative size of
this old population varies from dSph to dSph, as does the subsequent star
formation history.  Nevertheless, the presence of this old population has
led to the qualitative statement that star formation commenced in the Galactic
halo dSph galaxies at an epoch similar to that for the formation of the
Galactic halo globular clusters, regardless of the dSph's location in the
proto-Galactic halo.  However, 
if we are to increase our understanding of the processes that occurred
during the earliest stages of the evolution of the Galaxy's halo, we need 
quantitative results.  In particular, we seek a quantitative answer 
to the question
``How similar in age are the `first generation' stars that formed in the
various components of the Galactic halo?''.  The advent of the WFPC2 camera
onboard the Hubble Space Telescope has made attempting to answer this 
question feasible, and a number of relevant studies have recently appeared.

For example, Grillmair et al.\ (1998) have used HST/WFPC2 observations of
a field near the centre of the Draco dSph to produce a colour-magnitude (c-m)
diagram that reaches well below the main sequence turnoff in this dSph.  
They then use
these data to suggest that Draco is 1.6 $\pm$ 2.5 Gyr older than the 
metal-poor halo globular clusters M68 and M92.  This result contrasts 
with previous expectations, based on Draco's 
relatively red horizontal branch morphology, that this dSph would prove to be
somewhat younger than Galactic halo globular clusters of comparable 
metallicity.  It should be kept in mind, however, that the Grillmair et al.\
(1998) observations apply to a field region where the presence of significant 
abundance and possible age ranges complicate the interpretation.  Less
ambiguous results require studies of dSph star clusters, since star clusters 
are single age and abundance populations.

The two most luminous of the Galaxy's dSph companions, Fornax and Sagittarius,
possess their own globular cluster systems, and for both these dSphs there
are new results on the ages of their star clusters.  For example, 
Montegriffo et al.\ (1998) have shown that Terzan 8, the most metal-poor of
the four globular clusters associated with Sagittarius, has the same age
as the metal-poor Galactic halo clusters M55 and M68.  The precision
of this result, however, is limited to approximately $\pm$2 -- 3 Gyr
by the uncertainities in their ground-based c-m diagram.  
For the Fornax dSph, Buonanno et al.\ (1998) have used WFPC2
images to produce c-m diagrams that reach below the main sequence turnoff
for four of the five Fornax globular clusters.  They find that these four
Fornax globular clusters have identical ages to within $\pm$1 Gyr.  As for
Draco, the similarity of these cluster ages contrasts rather strongly with the 
expected age range of $\sim$2 Gyr based on the horizontal
branch morphology differences shown by the Fornax cluster c-m diagrams and the 
assumption that age is the ``second parameter''.  
As Buonanno et al.\  (1998) note, the result of closely similar ages for
all four clusters suggests that either horizontal
branch morphology is more sensitive to age than previously thought,
or some other quantity besides age is responsible for the horizontal branch 
morphology differences.  Buonanno et al.\ (1998) then go on to compare
their Fornax cluster c-m diagrams with those for Galactic halo globular
clusters.  They find ages that are not significantly different, at the 1 -- 2
Gyr level, from those for the Galactic halo clusters M92 and M68.

Thus, to a precision of $\sim$1 -- 2 Gyr, we can conclude that Fornax, Sgr and
Draco (and also probably Ursa Minor -- see Olszewski \& Aaronson 1985) did
indeed commence forming stars at the same time as the metal-poor globular 
clusters were forming in the proto-Galactic halo.  Other recent results
allow this conclusion to be widened.  In particular, 
Harris et al.\ (1997) have shown that the metal-poor
globular cluster NGC~2419, which lies in the extreme outer halo at a
Galactocentric distance of $\sim$90 kpc, has an age that is indistinguishable
from that of M92 to a precision of better than 1 Gyr.  Similarly, 
Olsen et al.\ (1998) and Johnson et al.\ (1998) have obtained WFPC2 data for a
total of eight globular clusters associated with the Large Magellanic Cloud.
Their results show that, first, to within an upper limit of $\sim$1 Gyr,
there is no detectable age range among these LMC star clusters.  Second, the
clusters are indistinguishable in age, at the $\sim\pm$1 Gyr level, from
Galactic halo globular clusters of comparable metal abundance.

All these results then suggest that, regardless of the subsequent star
formation histories, the {\it initial} epoch of star formation was well 
synchonized among {\it all} the components of the proto-Galactic halo, 
which may well have been distributed over a volume at least $\sim$100 kpc in radius.  In other words, despite the very different locations,
masses, densities and dark matter contents of the proto-LMC, the
proto-dSphs and the proto-NGC~2419 gas clouds, etc, in the proto-Galactic 
halo, the initial episode of star formation in all these components seems to
have been well co-ordinated.  An understanding of how this comes about would
undoubtedly advance our knowledge of galaxy formation and of conditions in the
early Universe.

\subsection{Abundance Ratios in dSph Red Giants}

The study of element abundance ratios, typically with respect to iron, in
the atmospheres of the members of a stellar system is important because such
ratios, and their variation with overall abundance, can provide significant
information on the enrichment processes that occur during the evolution of
the stellar system.  In particular, abundance ratio studies of stars in
dSph galaxies should be capable of providing direct constraints on their
chemical evolution.  It is also possible that such studies {\it might} provide
a signature to mark those Galactic halo field stars that have come from
disrupted dSph galaxies.  The determination of abundance ratios for
dSph stars is no easy task.  Even the brightest red giants in
the nearest dSphs are relatively faint and thus a large telescope is required.
The results that are described below come from the Keck telescope and the
HIRES spectrograph, but they should be regarded as precursors for what will
undoubtedly be an extensive area of study once other large telescopes (e.g.\
HET, Gemini, VLT, Subaru, Magellan, etc) begin science operations.

The first such study is that of Shetrone et al.\ (1998) who have analyzed
high dispersion spectra of four red giants in Draco.  They find, firstly,
that these stars show a substantial range in iron abundance: the [Fe/H]
values are --3.0, --2.4, --1.7 and --1.4 dex, respectively, 
where in each case the uncertainty 
in the [Fe/H] value is $\sim$0.1 dex.  The existence of this large
abundance range comes as no real surprise since we have known for some time
that most, if not all, Galactic halo dSphs possess significant internal
abundance ranges (e.g.\ Suntzeff 1993 and references therein).  However,
studies of large unbiased samples of red giants in dSphs from which to
determine the abundance distribution functions are generally lacking at the
present time.

The abundance ratios for these Draco red giants do, nevertheless, reveal
some interesting differences from globular cluster and field halo red
giants.  These are illustrated in Fig.\ 2 where the Shetrone et al.\ (1998)
results for Draco and for red giants in the globular clusters M92 
([Fe/H] = --2.27) and M3 ([Fe/H] = --1.53) are compared with the results
of McWilliam et al.\ (1995) and McWilliam (1998) for metal-poor field halo
stars.  For the $\alpha$--element calcium, the [Ca/Fe] values for the
globular cluster and field halo stars shown in the upper panel of Fig.\ 2
are consistent with the trends exhibited by larger samples of stars (see,
e.g., Norris, these proceedings).  However, the Draco stars, especially the
two more metal-poor objects, have [Ca/Fe] values that are significantly
lower than the overall trend.  These two stars have [Ca/Fe] $\approx$ 0.1 while
the 24 field halo stars with --3.2 $\le$ [Fe/H] $\le$ --2.0 in the McWilliam 
et al.\ (1995) sample have $<$[Ca/Fe]$>$ = 0.42 $\pm$ 0.02 dex.  On the
other hand, as the middle panel of Fig.\ 2 shows, the results for magnesium,
which is also an $\alpha$--element, show no such effect.  The two metal-poor
Draco red giants have [Mg/Fe] values consistent with the field halo star and
globular cluster red giant determinations.  The low [Mg/Fe] value for one
of the more metal-rich Draco red giants will be discussed below.

\begin{figure}
\begin{center}
\begin{minipage}{17.0cm}
\epsfxsize=8.23cm
\epsfbox {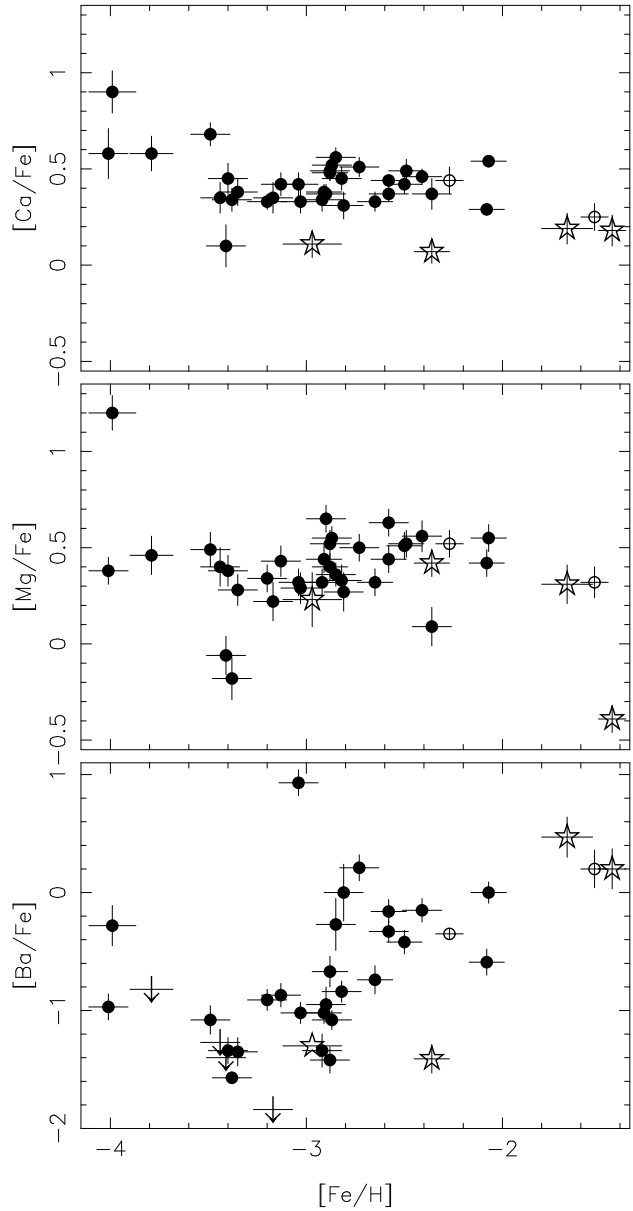}
\end{minipage}
\caption{Abundance ratios as a function of [Fe/H] for the $\alpha$--elements
Ca (upper) and Mg (middle) and for the s-process element Ba 
(lower).  In each panel filled symbols are field halo stars from
McWilliam et al.\ (1995) and McWilliam (1998).  The open circles are mean
values for 5 red giants in the globular cluster M92 and 6 in M3, while
star symbols represent individual red giants in the Draco dSph.  These data 
come from Shetrone et al.\ (1998).  In the lower panel the point for the
most metal-poor Draco star is an upper limit, not a measurement.  Other
upper limits are shown by downward arrow symbols.}
\end{center}
\end{figure}

The lower panel of Fig.\ 2 shows the results for the s-process element barium.
The two more metal-rich Draco giants have
[Ba/Fe] values that are consistent with the results of McWilliam (1998) for
a sample of metal-poor field halo stars.  This is also true for the 
globular cluster red giants.  However, as was found for [Ca/Fe], the lower 
panel of Fig.\ 2 reveals that the two more metal-poor
Draco red giants have significantly lower [Ba/Fe] values than do
field halo stars with similar [Fe/H]\@.  For the Draco star D24 this difference
in [Ba/Fe] is $\sim$1 dex while for Draco star D119 the difference is 
indeterminate since there is only an upper limit on [Ba/Fe] for this
most metal-poor star.
This upper limit though corresponds to the lowest measured values of [Ba/Fe]
in the McWilliam (1998) sample.  What are we to make of these results?  If
they are substantiated by a larger sample of stars observed at higher S/N
(the Shetrone et al.\ 1998 Draco spectra have S/N $\approx$ 24--30), then
they might well indicate that the IMF in the proto-Draco gas cloud was 
different from that in the Galactic halo.

One further point deserves comment.  As the middle panel of Fig.\ 2 shows,
the most metal-rich of the Draco red giants studied by Shetrone et al.\ (1998)
exhibits a significant Mg depletion.  As Shetrone et al.\ (1998) point out,
this star also possesses an oxygen depletion and a modest enhancement of
sodium.  Together with a postulated enhancement of aluminium 
(Al was not observed),
these abundance anomalies are reminiscent of the correlated CNO/NaMgAl 
abundance variations that are observed among the red giants in many globular 
clusters (e.g.\ Da~Costa 1997b and references therein).  The origin of these
abundance anomalies remains uncertain but in this context we need only note 
that the phenomenon is restricted to globular cluster red giants; 
it is virtually unknown among field halo red giants.  Consequently, {\it if} 
at least approximately 1 in 4 Draco red giants show these abundance anomalies, 
and {\it if} this fraction is typical for all Galactic dSphs, then the virtual
complete absence of such anomalies in field halo red giants suggests that
disrupted dSphs (or disrupted globular clusters for that matter) did {\it not} 
contribute significantly to the field halo population, contrary to the
suggestions of some other contributions at this meeting.  Clearly, a full
accounting of the frequency of occurrence of CNO/NaMgAl abundance anomalies
among dSph red giants is urgently needed.

A second high dispersion study of abundances and abundance ratios 
in Galactic dSph red giants 
is that of Smecker-Hane et al.\ (1998) for stars in the Sagittarius dSph.
This dSph is known to have a large internal abundance range.  For example,
the most metal-poor of the four 
Sgr globular clusters, Ter 8, has [Fe/H] $\approx$
--2.0 while the most metal-rich, Ter 7, has [Fe/H] $\approx$ --0.5 (e.g.\
Da~Costa \& Armandroff 1995).  The Smecker-Hane et al.\ (1998) results for
individual Sgr red giants are based on Keck + HIRES spectra that have
S/N $\sim$ 80.  They find, for the first three stars in their sample analyzed,
[Fe/H] values of --1.30, --1.03 and +0.11 dex.  This last value is
remarkably high; for example, it is significantly larger than the present-day
abundance, [Fe/H] $\approx$ --0.3, in the LMC\@!  Yet, this star, once Sgr is
fully disrupted, will be become a ``field'' object in the Galactic halo.

Of particular interest here though are the abundance ratio results.  For the
Sgr red giant with [Fe/H] $\approx$ --1.30, the [$\alpha$/Fe] ratios have
values of $\sim$0.3 dex, which are perfectly consistent with the ratios
observed in globular cluster red giants and in field halo stars (cf.\ Fig.\ 2
and Norris, these proceedings).  There is therefore nothing particularly
remarkable about this Sgr red giant.  For the Sgr red giant with [Fe/H] 
$\approx$ +0.11, however, the [$\alpha$/Fe] ratios are indeed noteworthy.
Smecker-Hane et al.\ (1998) find [O/Fe] $\approx$ --0.41, [Ca/Fe] $\approx$ 
--0.24, [Si/Fe] $\approx$ +0.06 and [Mg/Fe] $\approx$ +0.11 so that overall,
[$\alpha$/Fe] $\approx$ --0.12 dex.  It is important to note that the low
[O/Fe] in this star is {\it not} the result of the CNO/NaMgAl abundance
anomaly effect seen in globular cluster red giants and in at least one Draco
star.  If the low [O/Fe] was due to this effect, then the abundances of sodium
and aluminium should be significantly enhanced, and that is not observed
in this Sgr star (Smecker-Hane et al.\ 1998).  Instead, it seems reasonable
to suggest that most of the iron in this star comes from Type Ia supernovae
rather than Type II, and that consequently, since the timescale for SNIa 
exceeds that of SNII, this star is part of Sgr's younger population.  Further, 
as Gilmore \& Wyse (1991) have shown, abundance
ratios of this type are expected when the star formation is {\it episodic}
rather than relatively continuous.  In essence, the long intervals of 
quiescence between
periods of star formation allow the iron abundance to build up via SNIa, while 
since there is no star formation in these intervals, no SNII occur to 
produce the $\alpha$-elements.  Consequently, while recognizing that these 
results come from a preliminary analysis for a single star, the element 
abundance ratios nevertheless suggest that Sgr has 
had an episodic star formation history.
At least qualitatively, this result is remarkedly consistent with those
presented by Mighell et al.\ (these proceedings).  Their HST/WFPC2 
colour-magnitude diagram based study independently suggests that Sgr has indeed
had an episodic star formation history.  The concurrence of these results then
indicates that we are moving towards a more complete understanding of the
necessarily entwined star formation and chemical enrichment processes that
occurred in this dwarf galaxy.

\section{Summary}

When considering the properties of the dSph galaxies present in the Galactic 
halo today, we must keep in mind that these galaxies have survived for a
Hubble time.  Consequently, if there was a large population of dSph-like
objects early in the life of the Galaxy which has now been mostly disrupted,
the dSph galaxies that we can observe at the present epoch may not have been
``typical'' members of this hypothesized early population.  In particular,
the current dSph galaxies may have orbits that make them less susceptible
to tidal disruption, and/or perhaps they have more massive and/or denser
dark matter halos.  Consequently, we shouldn't necessarily expect exact 
correspondence between halo properties and those of present-day dSphs even if
the Galactic halo does have a substantial contribution from disrupted dSphs.
Nevertheless, the present-day dSphs are intriguing objects for further study
both:\\
{\it dynamically}, since we are beginning to see the emergence of extensive
observational datasets and more complex theoretical models of both the
dSphs and their interaction with the Milky Way; and, from a\\
{\it stellar populations} point-of-view, where increased knowledge of star
formation histories, and abundance and abundance ratio distributions will
tell us a lot about the evolution of these lowest luminosity galaxies.

\acknowledgements

I would like to thank Ed Olszewski for a timely message indicating the 
availability of a preprint, Tammy Smecker-Hane for making
available results from her high dispersion studies of Sgr red giants
in advance of publication, and John Norris for his comments on a draft 
of this paper.  I would also like to place on record
my indebtness to the late Prof.\ Alex Rodgers, to whom this meeting is
dedicated.  Our discussions in recent years ranged 
from tenure ratios in the IAS and large telescopes in Australia to the globular
clusters associated with Sgr dSph; I always found the discussions beneficial.
Mt.\ Stromlo \& Siding Spring Observatories, and Australian astronomy in
general, are poorer for his untimely demise.

\end{document}